\documentclass[aip,jap,reprint,amsmath,amssymb,citeautoscript]{revtex4-1}

\usepackage{graphicx}
\usepackage{dcolumn}
\usepackage{bm}
\usepackage{natbib}

\usepackage[version=3]{mhchem} 
\usepackage{amsmath}

\begin{document}
\preprint{AIP/123-QED}
\title{Towards the Ionizing Radiation Induced Bond Dissociation Mechanism in Oxygen, Water, Guanine and DNA Fragmentation: A Density Functional Theory Simulation}

\author{Santosh KC}
\affiliation{Chemical and Materials Engineering, San José State University, San José, CA 95192, USA.}
\author{Ramin Abolfath}
\affiliation{Department of Radiation Physics, University of Texas MD Anderson Cancer Center, Houston, TX, 75031 USA.}
\date{\today}
\begin{abstract}
The radiation-induced damages in bio-molecules are ubiquitous processes in radiotherapy, radio-biology and critical to space-projects.
In this study we present a precise quantification of the fragmentation mechanisms of
deoxyribonucleic acid (DNA) and the molecules surrounding DNA such as oxygen and water under non-equilibrium conditions using the first-principle calculations based on density functional theory (DFT).
Our results reveal the structural stability of DNA-bases and backbone that withstand up to a combined threshold of charge and hydrogen abstraction owing to simultaneous direct and indirect ionization processes.
We show the hydrogen contents of the molecules significantly control the stability in the presence of radiation.
This study provides comprehensive information on the impact of the direct and indirect induced bond dissociations and DNA damage, and introduces a systematic methodology in fine-tuning of the input parameters necessary for the large-scale Monte Carlo simulations of radio-biological responses and mitigation of detrimental effects of ionizing radiation.
\end{abstract}
\maketitle

\section{Introduction}
Deoxyribonucleic acid (DNA), is one of the key components of life which is responsible for the storage and transmission of genetic information \cite{Watson1953, Dekker2001}. It comprises a phosphate backbone and four nitrogen-containing bases, which are named adenine (A), cytosine (C), guanine (G), and thymine (T). It is found that A base pairs only with the T base, and the G base pairs with the C base  \cite{Watson1953, Dekker2001}. DNA is an important part of life, but, it is very sensitive to the environment.

In radiotherapy, the interaction of Mega-voltage ionizing radiation with biological systems causes ionization processes in biomolecules such as DNA, proteins, and their surrounding environment in cell nuclei. Among all of these ionization processes, DNA damage is critical to the clinical outcome of radiotherapy.
After initial induction of DNA damage, a dynamical cascade of stochastic microscopic events and complex biochemical pathways, including, enzymatic homologous and non-homologous repair and misrepair end-joining determine the lethality of the irradiated cells.

Empirical studies in radio-biology and radio-chemistry have suggested induction of approximately 1000 single-strand breaks (SSBs) and 40 double-strand breaks (DSBs) per one gray (1Gy = 1 J/kg ) of low linear energy transfer (LET) of ionizing radiation such as X- or $\gamma$-rays in typical mammalian cells
[\onlinecite{Ward1988:NARMB,Goodhead1994:IJRB,Nikjoo1997:IJRB,Semenenko2004:RR}].

Accordingly, the level of DNA molecular base damage has been estimated to be around 2,500 to 25,000 per gray in a cell, which is about 2.5-25 times the yield of sugar-phosphate-induced damage in the DNA backbone.

The occurrence of initial DNA damage has been classified into direct and indirect processes.
In direct mechanism, ionization takes place via direct electrodynamical coupling between the source of radiation
and DNA molecule.
For X- or $\gamma$-rays, depends on the energy of the incident photon, the coupling varies among photoelectric and Compton effects where shell electrons are ejected directly.
In addition, high enough energy photons may interact with the nuclei of atoms and generate pair of electrons and positrons. Another source of uncharged particles such as neutrons may undergo nuclear interaction and make nuclear fragmentation and produce secondary charged particles as well as photons.
The charged particles, either primary or secondary, interact with shell electrons through long-range Coulomb interaction. Under enough energy and momentum transfer, these charged particles eject shell electrons.
Thus a ``direct damage" originates from the direct ionization of a molecule, i.e., an isolated molecule (in vacuum) loses a number of shell electrons within atto-second (electromagnetic) time-scales and undergoes structural instability because of electrostatic charge imbalance and the repulsive forces among positively charged nuclei.
The threshold of such instabilities requires a minimum number of ionizations and energy on a specific site of DNA.

In the indirect mechanism of radiation interactions, the radiation dominantly ionizes water molecules and creates neutral \ce{^{.}OH} free radicals [\onlinecite{Ward1988:NARMB}].
The DNA damage process involves the generation and diffusion of \ce{^{.}OH} radicals
in cell nuclei and/or aqueous environments followed by chemical reactions that allow the removal of hydrogen atoms from the DNA. This process is energetically favorable for \ce{^{.}OH} radicals as it forms a water molecule and fills the electronic shell by neutralizing its magnetic moment.

On the other fronts, there is a tremendous concern about the risk of radiation in the human body while going into outer-space \cite{Hellweg2007}. Outer-space consists of an ionizing radiation environment dominated by energetic and penetrating ions and nuclei. Thus, there is a risk of DNA damage due to ionizing radiation and a chance of radiation-induced cancer in manned space exploration \cite{Hellweg2007,NCRP2002}.
Like in radiotherapy and radiobiology, there is a need for atomic-level understanding of biomolecules in radiation exposed in space.
A large-scale computational model, relying on quantum dataset, will provide more realistic computational tools in assessing the  biological risks due to space radiation, in particular for astronauts who are planning for the long-term exploration of other planets such as Mars.
This is in alignment with NASA's space radiobiology research that aims to mitigate the detrimental effects of the space radiation environment on the human body, a project focusing on the human presence outside of the relative protective Van Allen belt.
Although the spacecraft itself somewhat reduces radiation exposure, it does not completely shield astronauts from galactic cosmic rays, which are highly energetic heavy ions, or from solar energetic particles, which primarily are energetic protons. By one NASA estimate, for each year that astronauts spend in deep space, about one-third of their DNA will be hit directly by heavy ions \cite{Chiara2016, Simonsen2020} from Galactic Cosmic Radiation (GCR).

We note that because of the aquatic environment in cells, approximately 70 to 80$\%$ of interactions take place through indirect damage and the rest are associated with the direct damage.
In recent years, various types of molecular simulations were devoted to studying DNA damage by either free radicals, or direct damage [\onlinecite{Abolfath2009:JPCB,Abolfath2010:JCC,Abolfath2011:JPC,Abolfath2012:JPCA,Abolfath2012:JPCA}].
Here we combine these two events to study their mutual effects.
Moreover, many current computational platforms designed for the large-scale simulations of the DNA-damage at the nanoscopic scales \cite{Agostinelli2003:NIMA,Incerti2010:IJMSSC,Schuemann2019:RR,Faddegon2020:PM,Friedland2011:MR,Plante2011:RPD,Lai2021:PMB}
lack accurate details from the first-principle direct and in-direct processes.
We aim to cover the gap in the details of the input parameters and allow the developers to update the tables used for MC simulation of DNA damage.

In this study, we focus on the simulation of combined direct and indirect damage to DNA molecules including base and backbone.
As a representative of DNA-base, and without loss of generality, we focus on Guanine. We find as a combined function of ionization and hydrogen loss in indirect mechanism, electrostatic repulsion of atomic nuclei dominates the electronic chemical bonds and molecular fragmentation takes place.
Thus we quantify DNA fragmentation as a function of ionization and hydrogen abstraction.
We show that at least four to five ionization must take place till the molecule undergo mechanical instability and fall apart.
Because in energy transfer by a high-energy photon or a charged particle adequate energy may transfer locally to DNA base or backbone, the such number of ionization can be scored.

The remainder of the paper is organized as follows. Section II introduces the calculation methods. The results and discussion of the results are described in Sec. III. Section IV provides the conclusion of the research.

\section{Computational Methods}
First-principles calculations based on Density Functional Theory (DFT)\cite{hohengerg1964,khonsham1965} are performed to investigate the charged defects in O$_{2}$, guanine, and DNA. The core and valence electrons interactions were described within projector-augmented plane-wave (PAW) potentials as implemented in the Vienna \textit{Ab-initio} Simulation Package (VASP)\cite{Kresse1996,Hafner1993,KF1996}. The exchange potential with the generalized gradient approximation of Perdew, Burke, and Ernzerhof (PBE) \cite{Perdew1996}. An energy cutoff of 500 eV is used for the plane-wave basis set in all the calculations. Spin polarization was used in all the calculations. In the PAW potentials used the 2s$^{2}$ 2p$^{4}$, 3s$^{2}$ 3p$^{3}$, 2s$^{2}$ 2p$^{2}$, 1s$^{1}$ and 2s$^{2}$ 2p$^{3}$ electrons were explicitly treated as the valence electrons for O, P, C, H, and N, respectively.
First, the  Oxygen and water molecule in a simulation box of size $15\AA \times 15\AA \times 15\AA$ was optimized. Electrons are gradually removed from the system to observe the oxygen and water bond dissociation. Similarly, we have investigated the effect of electron extraction in Guanine and DNA molecules. The charged molecules were relaxed until the Hellman-Feynman forces were less than 0.01 eV/\AA . There have been successful reports of using DFT-based computational approach with plane-wave basis set on Guanine and DNA \cite{Felice2001,Errez2008}.

\begin{figure}
\includegraphics[scale=.4]{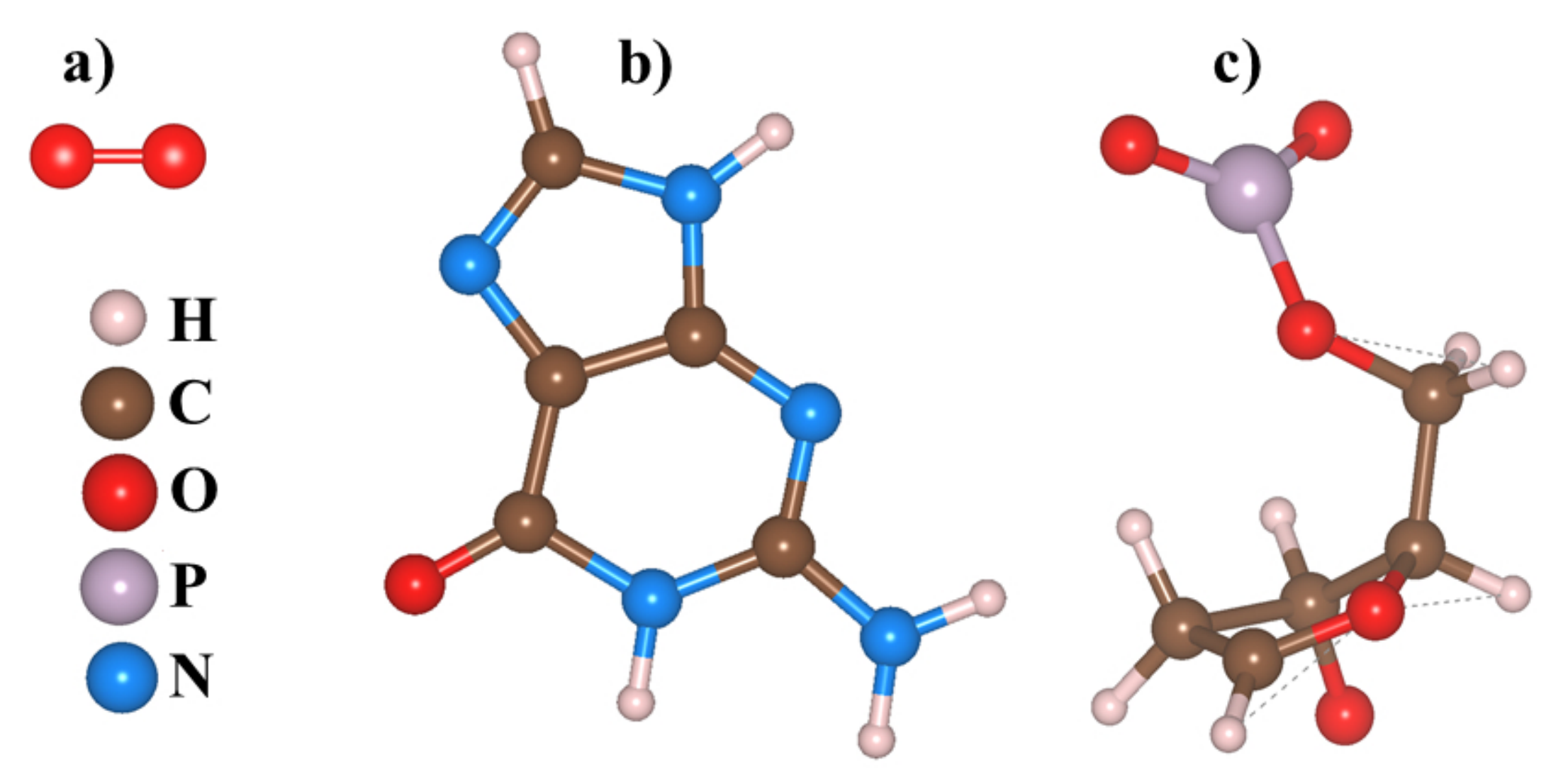}
\caption{\label{fig:1} Optimized atomic structures of the Oxygen, Guanine (C$_{5}$H$_{5}$N$_{5}$O), and fragment of Deoxyribonucleic acid (DNA). a) Oxygen b) Guanine  and c) fragment of DNA}.
\end{figure}

\section{Results and Discussion}
Our DNA backbone model is similar to the deoxyribose residue used in other {\it ab-initio} calculations such as Ref.~[\onlinecite{Miaskiewicz1994:JACS}] where the system of interest consists of a DNA nucleobase modeled by an amino group attached to the deoxyribose as shown in Figure~\ref{fig:1}(c).
The deoxyribose sugar ring is an important component of nucleotides and plays a role in the stability of DNA double-helix structure. Any damage to the sugar ring causes breakage of strands and the formation of single-strand break (SSB).  The formation of a pair of SSBs in opposite strands, within ten base pairs, leads to a single double-strand break (DSB).

To study the effect of ionizing radiation on the molecules, we systematically perform the DFT calculation of molecules in various charge states. For a representative of DNA-base, we consider Guanine in addition to the oxygen and water molecules in our simulations.


\subsection{Effect of electron extraction on oxygen molecules}

It is know that oxygen species play important roles in both tumor and normal cells. Typically, tumor cells contain less oxygen with a complex environment known as hypoxic, so they are more radio-resistant than normal tissues. Also, there are specific transitions in oxygen that make the molecular oxygen toxic, such as singlet oxygen. Thus, it is critical to understand how the charge induced radiation environment facilitates the dissociation of oxygen molecule.

\begin{figure}
\includegraphics[scale=.6, angle=0]{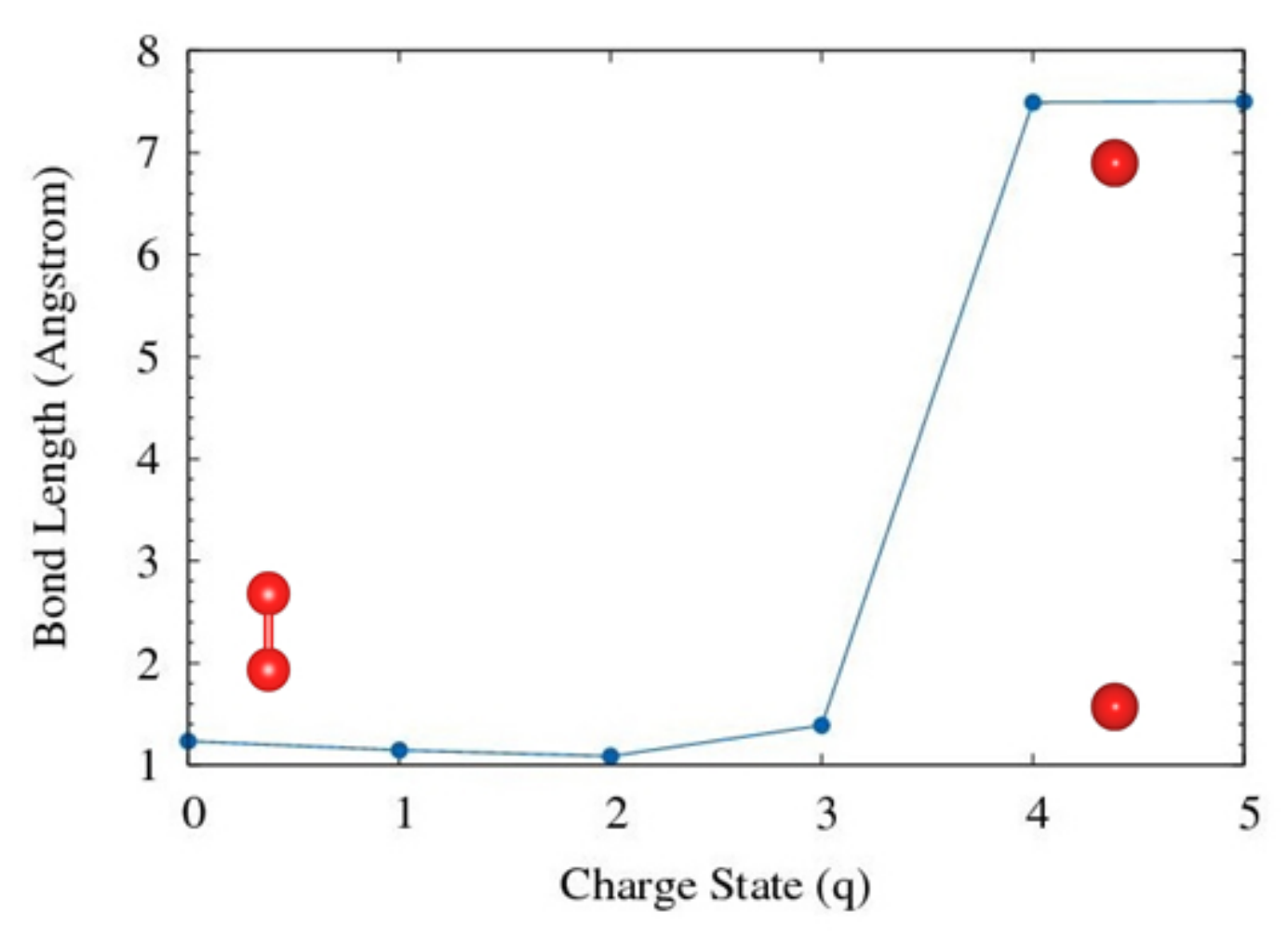}
\caption{\label{fig:2} The effect of electron removal in oxygen molecule. The dots represents the O-O bond,the relative bond distance is indicated by O atoms.}
\end{figure}

First, the oxygen (O$_{2}$) molecule was optimized and bond distance and equilibrium energy were obtained. The O-O bond length was found to be 1.233 \AA, which is consistent with experimental and previously reported computational values.
In O$_{2}$ molecule, we notice that gradual electron extraction shows initially the bond length contracts slightly then expands for when a large number of electrons are removed (See Figure ~\ref{fig:2}). The bond length variation as a function of the charge state of an oxygen molecule is presented in table ~\ref{tab:table1}. The removal of electrons weakens the bond strength and hence the bond-dissociation energy is reduced. In a pure oxygen molecule, the bond-dissociation energy is stronger due to the formation of double bonds (119 kcal/mole or 5.15 eV/bond).

\begin{table}[]
\caption{\label{tab:table1} The effect of electron removal in the bond length (d) of O$_{2}$ molecule. The charge state (q) refers to the number of electron removed from the neutral system. Thus, q=1 refers to 1.60217662 $\times 10^{-19}$ coulombs per molecule.}
\begin{tabular}{ccc}
\hline
Charge (q) & Bond Length d ($\AA$) & $\Delta d (\AA)$ \\
\hline
 0  & 1.233  & 0 \\
 +1  & 1.146  &-0.087 \\
 +2  & 1.085   &-0.148 \\
 +3  & 1.387  &+0.154 \\
 +4  &7.491 &+6.258\\
 +5  &7.500 &+6.267\\
 \hline
\end{tabular}
\end{table}

The figure shows the bond length as a function of the charge state (q).  The x-axis label 1 refers to the +1 charge state with removing one electron, resulting in a positive charge in the molecule. We observed that with q=+4 (4 electrons are removed) demonstrates the dissociation of the bonds. The distance (d=7.5 \AA) is due to chosen box size of 15 \AA, indicating that they are isolated from each other.

It is found that oxygen depletion leads to lower normal-tissue toxicity at FLASH dose rates that take place within femto-to nanoseconds of irradiation. The biomolecular damage would be reduced in an environment with physoxic oxygen levels \cite{ramin2020}.

\subsection{Effect of electron extraction in water molecules}

\begin{figure}
\includegraphics[scale=.34, angle=0]{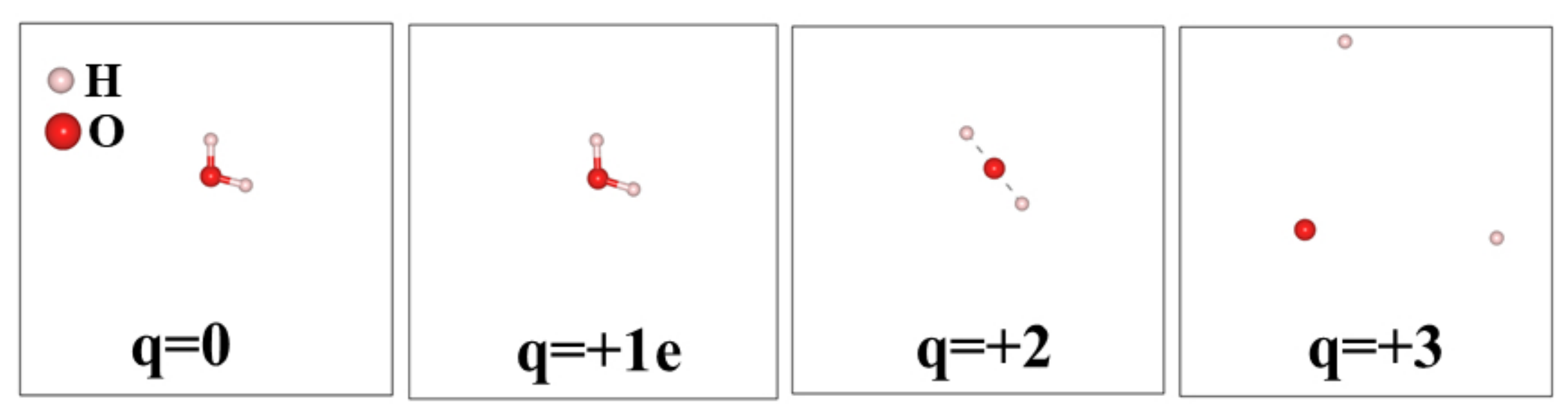}
\caption{\label{fig:3} The effect of electron removal in water molecule.}
\end{figure}

\begin{figure}
\includegraphics[scale=.65, angle=0]{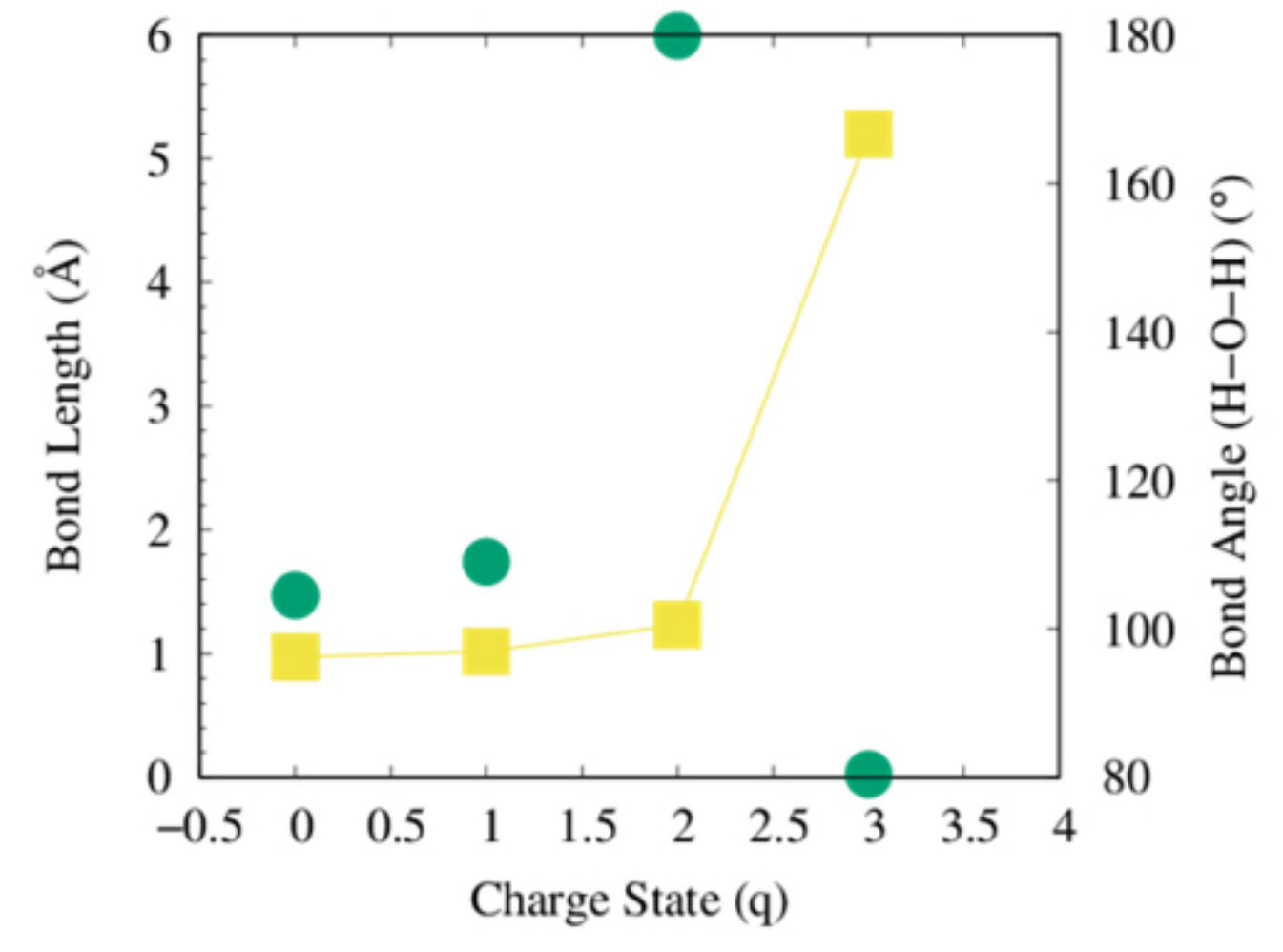}
\caption{\label{fig:4} The effect of electron removal in water molecule. The dots represents the H-O-H bond angles.}
\end{figure}

Water molecules (H$_{2}$O) which are ubiquitous and are a significant part of life processes, stabilized as a tri-atomic molecule with \textit{C2v} molecular symmetry and bond angle of 104.5\textdegree  between the oxygen atom and the two hydrogen atoms. The H-O bond length is close to
the bond (O–H) length of 0.9572 \AA and the bond angle (H–O–H) of 104.5 \textdegree. Our calculated data are in very good
agreement with the experimental reports [See Table ~\ref{tab:table2}. We observed that upon extraction of electrons from the water molecule, both the bond lengths (H-OH) and the bond angle (H-O-H) change significantly as shown in Fig.~\ref{fig:3} and  ~\ref{fig:4}. Moreover, we would like to mention that these calculations are performed in a vacuum. For example, the pathway of dissociation of the water molecule, surrounded by other water molecules would be significantly different. It stabilizes OH-radical. In Fig. 3, none of these scenarios correspond to the formation of OH-radical, simply because of symmetry and periodic boundary condition.

\begin{table}[]
\caption{\label{tab:table2} The effect of electron removal in the bond length of H$_{2}$O molecule. The charge state (q) refers to the number of electron removed from the neutral system. Charge q=1 refers to 1.60217662 $\times 10^{-19}$ coulombs per molecule. The H-O-H bond angle is also provided.}
\begin{tabular}{ccccc}
\hline
Charge (q) & Bond Length d ($\AA$) & $\Delta d (\AA)$ & $\Theta$ \textdegree & $\Delta \Theta$ \textdegree\\
\hline
 0  & 0.972 & 0     & 104.502 & 0 \\
 +1 & 1.017 & +0.045 & 109.004 & +4.502\\
 +2 & 1.232 & +0.260 & 179.862 & +75.360\\
 +3 & 5.200 & +4.228 & 80.416  & -24.086\\
\hline
\end{tabular}
\end{table}

The dissociation of the HO-H bond in a water molecule needs approximately 118.8 kcal/mol (497.1 kJ/mol) when there is no charge is involved. The bond energy of the covalent O-H bonds of the water molecule is approximately 110.3 kcal/mol (461.5 kJ/mol)\cite{Lehninger2005}. In the case of the ionizing environment, these values will be reduced and hence, this reduction causes the fragmentation of the bonds easily as shown in Fig ~\ref{fig:3}). Figure ~\ref{fig:4} shows how the bond length and the angles changes with the application of increasing charge (removal of an electron from the water molecule).

For a highly ionizing environment, the bond angles will deviate from the angular to planar before breaking the bonds as shown in the case of charge state (q=2).
This indicates that the 2 electrons extraction per water molecule (3.20435324$\times 10^{-19}$ coulombs per molecule) is sufficient to drive the fragmentation into ions.

\subsection{Effect of electron extraction in guanine and in a fragment of DNA molecules}

Finally, the impact of the electron extraction in guanine and a fragment of DNA molecule is studied. Guanine is one of the four main nucleobases that exist in DNA. Guanine (2-Amino-1,9-dihydro-6H-purin-6-one: IUPAC) consists of a fused pyrimidine-imidazole ring system with conjugated double bonds and has a planar molecular structure.
To avoid spurious interaction due to the periodic boundary condition (PBC) in DFT calculation, a large simulation box was adopted for each molecule.
Since, these molecules have multiple bonds, instead of monitoring individual bond length, we note the sum of the atomic displacements compared to the initial configurations. The gradual fragmentation of the Guanine molecule is observed as shown in Fig. ~ref{fig:6}. The corresponding sum of the displacements of the atoms as a function of charge states is shown in Figure ~\ref{fig:7}. We noticed that when the charge state is 4e, the C-C double bond is broken that will drive the structural instability.

\begin{figure}
\includegraphics[scale=.3, angle=0]{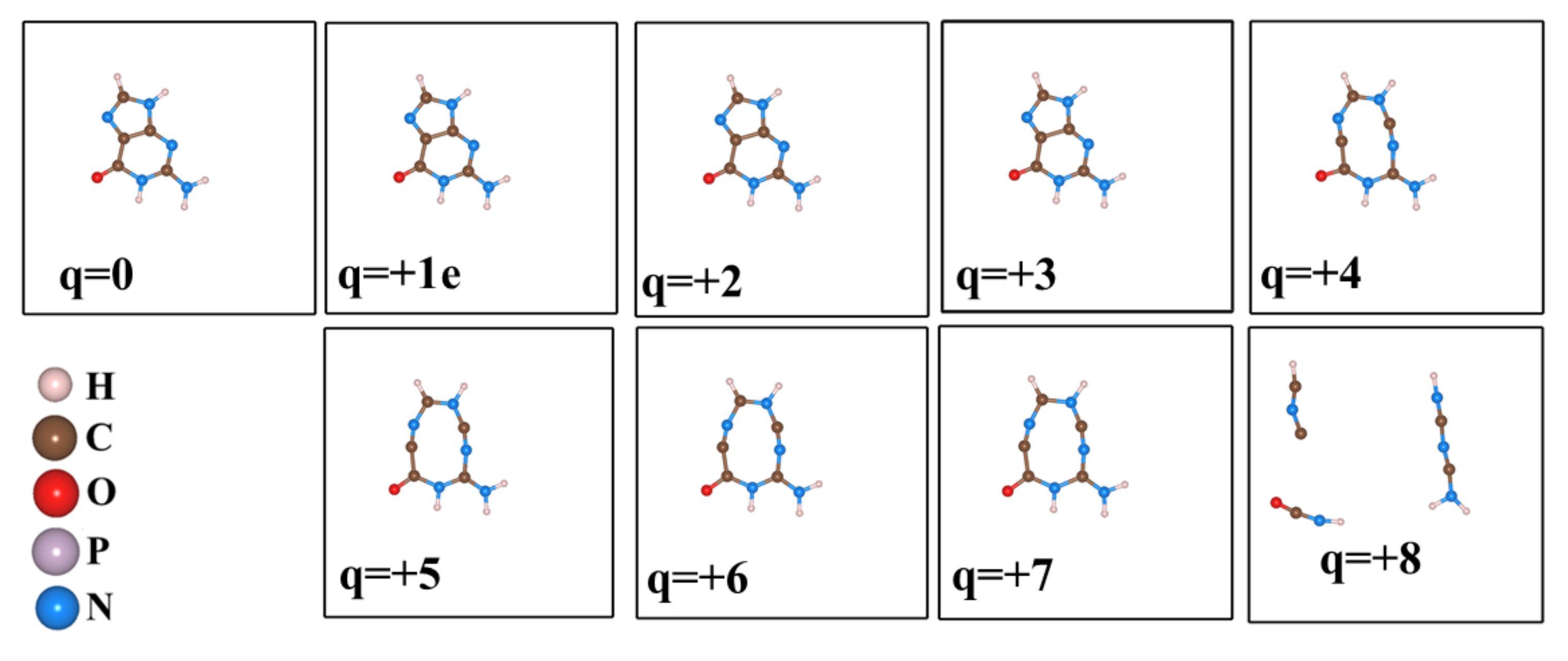}
\caption{\label{fig:5} The effect of electron removal in Guanine molecule. }
\end{figure}

\begin{figure}
\includegraphics[scale=.5]{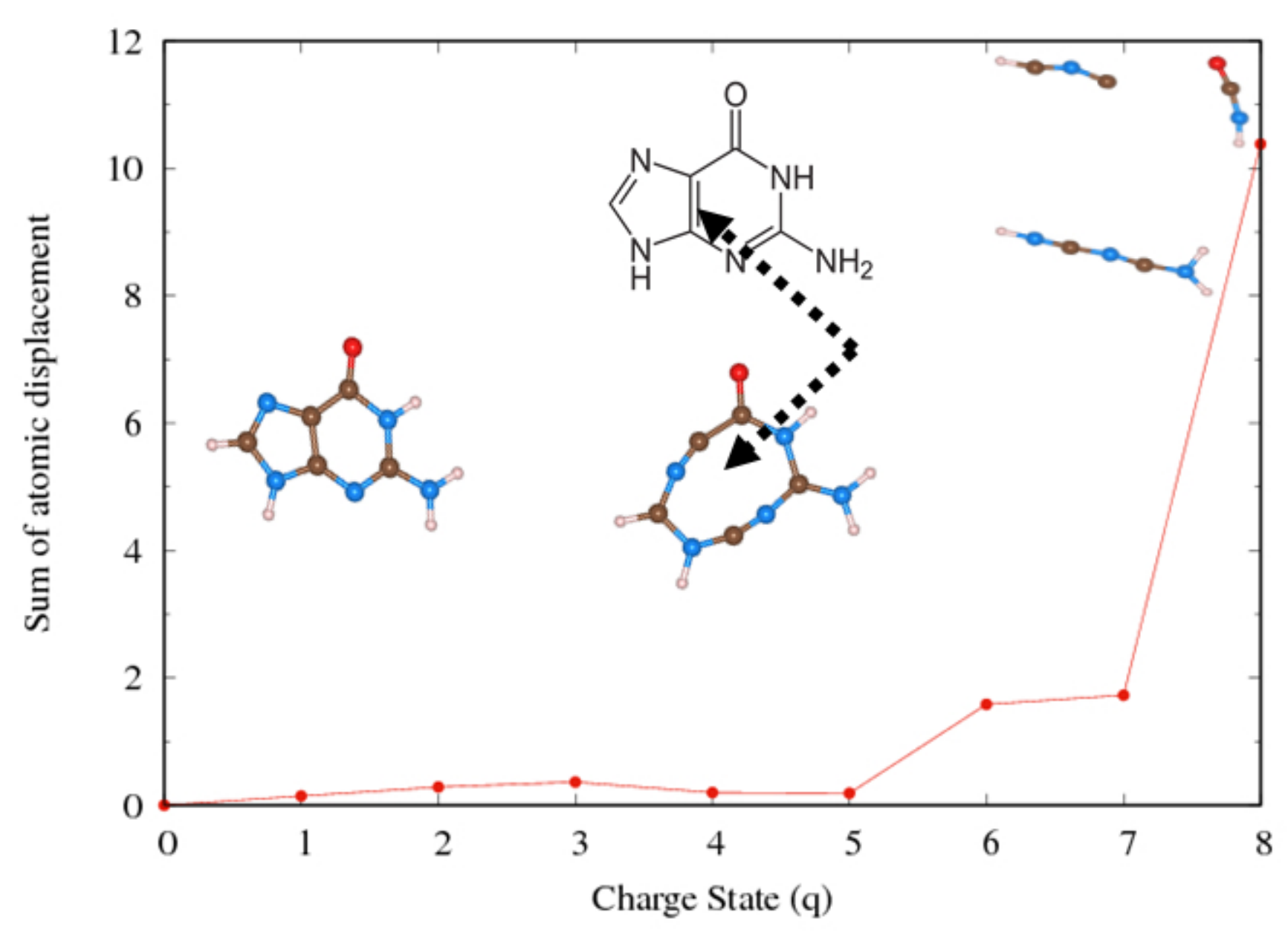}
\caption{\label{fig:6} The sum of the displacement of the atoms as a function of charge (q) in a Guanine. The sum of the displacement is in \AA and charge is in terms of number of electrons removed.}
\end{figure}

\begin{figure} [h]
\includegraphics[scale=.31, angle=0]{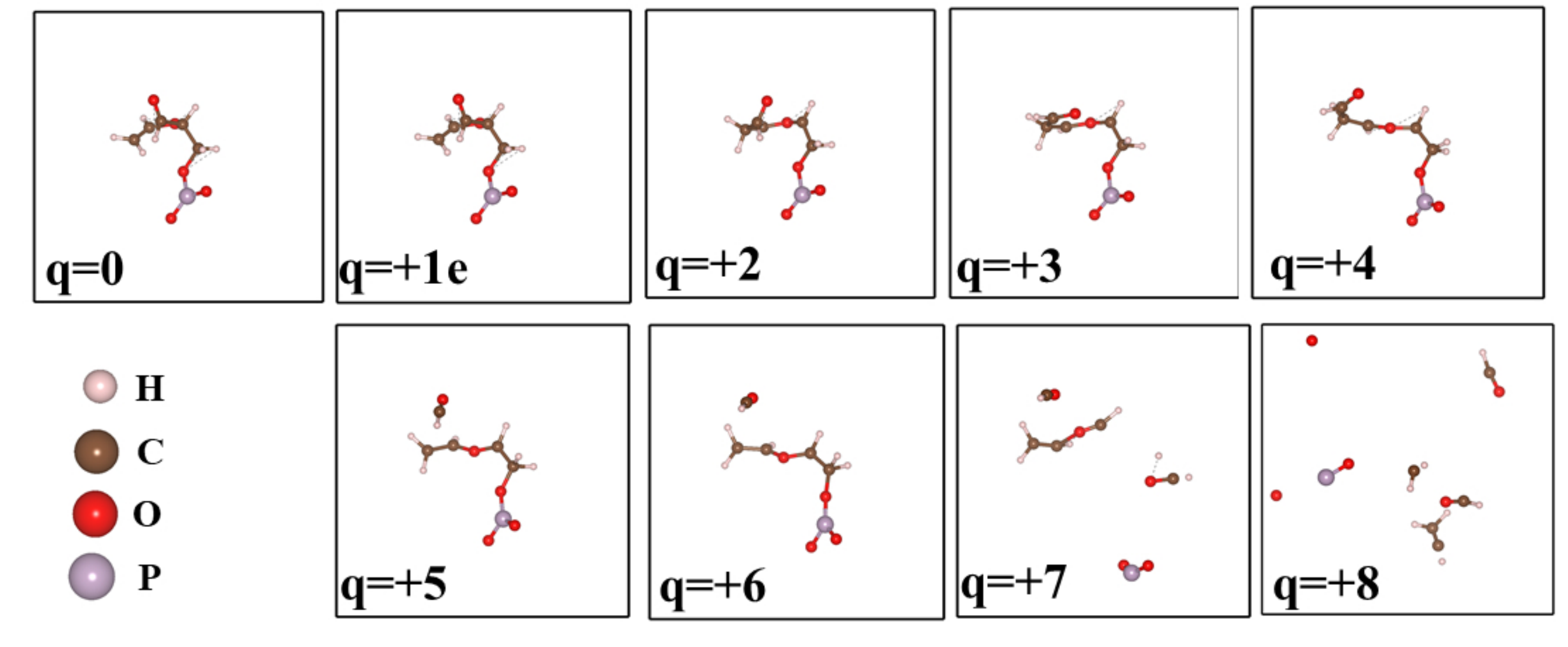}
\caption{\label{fig:7} The effect of electron removal in a fragment of DNA molecule. }
\end{figure}

Similarly, upon extraction of electrons from a fragment of DNA  as shown in figure ~\ref{fig:7}, bonds start to change and dissociate when the change is sufficient such as q=4e. Eventually, the molecule starts to collapse into smaller fragments in a sufficiently high ionization environment. This indicates that these molecules are prone to damage when exposed to an ionizing radiation environment.
In addition, the temperature and pressure might also play a role in altering this behavior to some extent. Thus, any chances of ionizing radiation-induced DNA damage should be checked seriously during radiation therapy treatment.

Ionizing radiation can extract electrons from these molecules resulting in ions that can trigger bond dissociation. Our results indicate that radiation directly affects DNA atomic structure by causing fragmentation. In addition, there might be secondary effects such as the creation of reactive oxygen species that oxidize proteins and lipids, and cause damages to DNA, eventually, the overall effect might cause cell death and mitotic catastrophe \cite{Soto2015}.

\subsection{Effect of hydrogen Contents in the electron extraction and stability of DNA}

\begin{figure}
\includegraphics[scale=.35, angle=0]{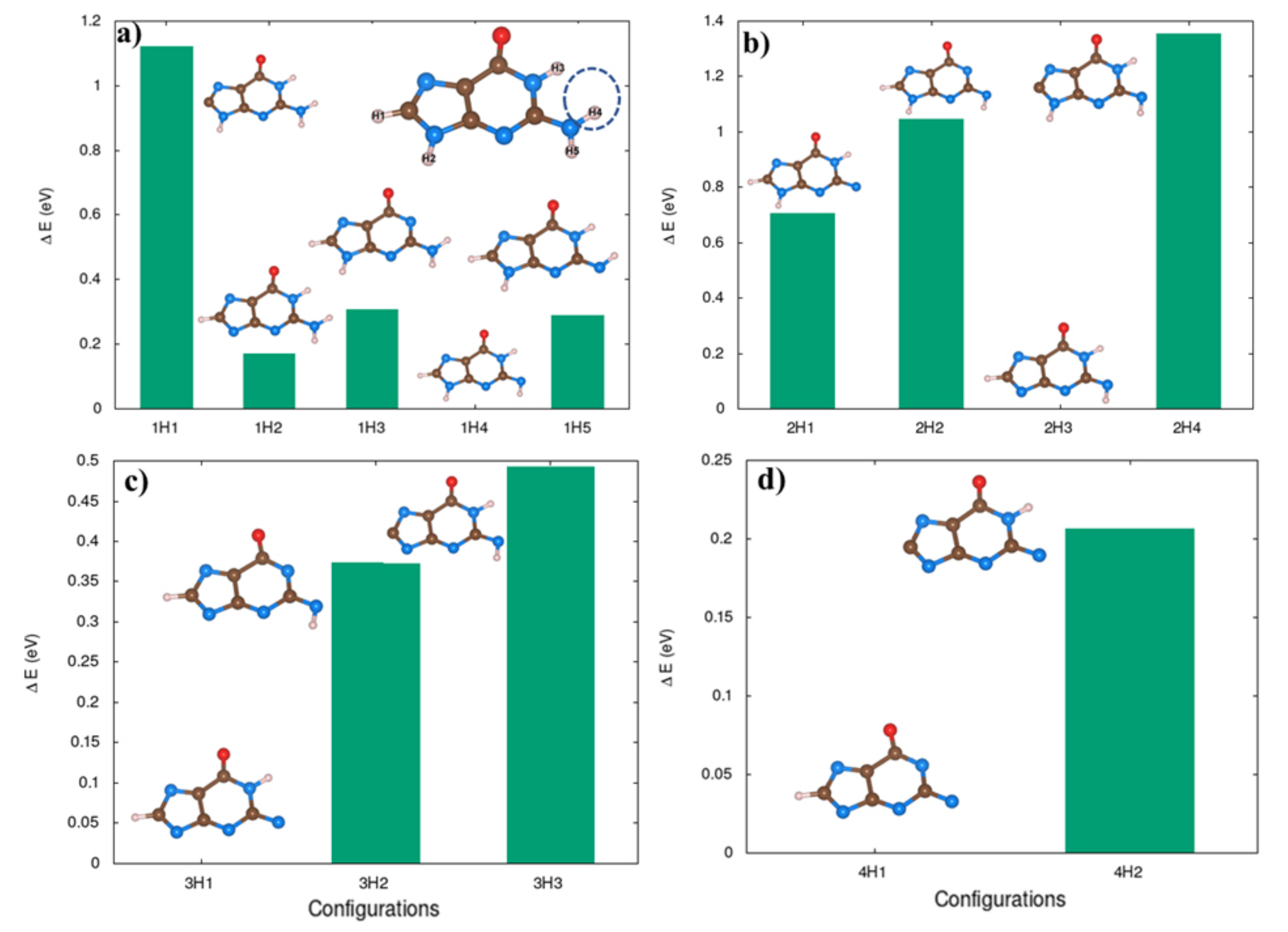}
\caption{\label{fig:8} The effect of hydrogen removal in Guanine molecule. Notation- 1H1 refers to one H removed the first configuration, 1H5 refers to 1H removed and is the fifth configuration. Similarly, 2H1: 2H atom removed and is the first configuration considered and so on. The energy of the configuration is with respect to the energetically most stable configuration in each H contents.}
\end{figure}

Moreover, hydrogen deficient molecules were also investigated in order to check their dependence on the charge-induced dissociation. As H atoms were gradually removed from Guanine as shown in ~\ref{fig:8}, it is observed that even with less electron removal can trigger the fragmentation of the molecules. Guanine cyclic ring is stable up to charge state of 3e for when up to 3H atoms are removed. However, they will be significantly modified when 4 or 5H atoms are removed. In 5H deficient case, the charge state of 3e completely dissociates the molecule into the molecular chain.

\begin{figure}
\includegraphics[scale=.31, angle=0]{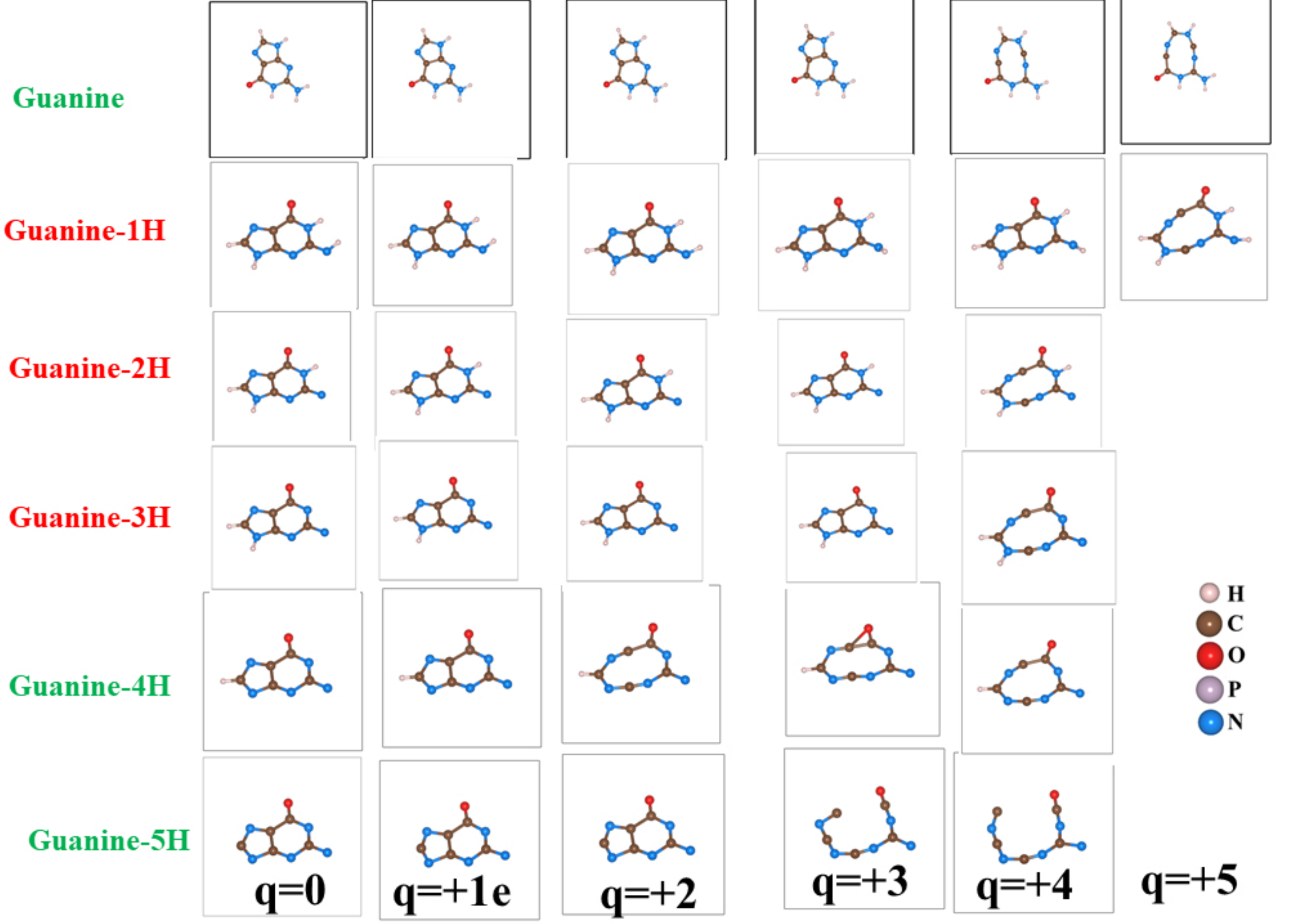}
\caption{\label{fig:8} The effect of electron extraction in hydrogen reduced Guanine molecule.}
\end{figure}

In addition, we observed that the fragment of DNA molecule dissociates upon removal of H atoms. The combined effect of charge and reduced H environment will lead to the fragmentation of these molecules. This indicates that the charge-induced dissociation of molecules strongly depends on the hydrogen environment.
\\
We optimize the geometry after the removal of H atoms, this may end up with a slightly different initial geometry for the removal of electrons. This optimized geometry would be different if we remove the electron first. Therefore, these two operations are not identical. We believe that this may happen in the actual scenario of DNA damage as the H-abstraction is a much slower process compared to direct damage that is electron removal. This is the case of damage induced by a single track of radiation that is responsible for alpha in the linear-quadratic cell survival model. The second scenario that is the removal of H and a subsequent direct ionization is most likely relevant to DNA damage induced by two tracks that is relevant to the beta term in the linear-quadratic cell survival model.

The current Monte Carlo (MC) codes utilized in studying the impact of radiation on biological materials lack the details since they use some empirical values for the excitations and DNA damage. This first-principles-based calculation provides important input parameters to take into account in those models.


\section{Conclusion}
Using DFT calculations, we systematically investigated the atomic bond dissociation in an ionization environment and the fragmentation behavior of the DNA base pair molecule along with water and oxygen molecules. Our results demonstrate that the bond fragmentation is proportional to the charge of the molecule and there is the limitation of the charge density of the molecule that it can withstand before collapsing into its fragments. This highlights the importance of using the optimal dose of radiation for safe use. Moreover, the bond dissociation behavior strongly depends on the hydrogen contents of the molecule. A hydrogen-reduced environment is detrimental to radiation-induced molecule fragmentation. This research is very applicable in radiation therapy as well as an environment where the human body will be exposed to radiation environments such as nuclear power plants or voyage to outer space. Thus, this study shed light on the atomic-level details of the mechanism of bond dissociation in the presence of ionizing radiation.

{\bf Acknowledgments:} S. KC acknowledges the faculty start-up grant provided by the Davidson College of Engineering at San José State University. We acknowledge the computational resources provided by the Extreme Science and Engineering Discovery Environment (XSEDE), which is supported by National Science Foundation grant number ACI-1548562 and the National Energy Research Scientific Computing Center (NERSC), a U.S. Department of Energy Office of Science User Facility operated under Contract No. DE-AC02-05CH11231.


\appendix

\end{document}